\documentclass[aps,twocolumn,showpacs, showkeys,nofootinbib,floatfix]{revtex4-1}
\usepackage{amsmath}
\usepackage{amsfonts}
\usepackage{txfonts}
\usepackage{graphicx}
\usepackage{dcolumn}
\usepackage{bbm}
\usepackage{amssymb}
\usepackage{latexsym}
\usepackage{titlesec}
\usepackage[colorlinks=true, linkcolor=red, citecolor=blue]{hyperref}
\usepackage{longtable}

\begin{document}

\title{Probing the Neutrino Mass through the Cross Correlation between the Rees-Sciama Effect and Weak Lensing}

\author{Lixin Xu}
\email{Corresponding author: lxxu@dlut.edu.cn}

\affiliation{Institute of Theoretical Physics, School of Physics \&
Optoelectronic Technology, Dalian University of Technology, Dalian,
116024, P. R. China}

\begin{abstract}
Cosmology plays a fundamental role to determine the neutrino mass, therefore also to determine its mass hierarchy, since the massive neutrino contributes to the total matter density in the Universe at the background and perturbation levels, once it becomes non-relativistic. After the non-relativistic transition the fluctuations are smashed out at the scales $k\gg k_{fs}$. Therefore, the missing fluctuation in the total matter is imprinted on the large scale structure, say the suppression of the matter power spectrum $\Delta P/P\approx -8f_{\nu}$ at the scales $k\gg k_{fs}$. In this paper, instead of considering the linear perturbation theory, which is well understood in the presence of neutrino, we propose to use the cross correlation between the Rees-Sciama effect and weak lensing to probe the neutrino mass. At the small scales, the density contrast grows faster than the background scale factor $\delta\sim a$, that makes a sign flipping on $\Phi' \propto \mathcal{H}\delta d\ln (\delta/a)/d\ln a$, which happens only in the non-linear regime. We show that the flipping scale in the cross power spectrum between the Rees-Sciama effect and weak lensing depends on the neutrino mass by assuming the shallow and deep weak lensing surveys. Our analysis shows that the Deep survey has larger signal-to-noise ratio $S/N\sim 160$. Finally, we use the Fisher information matrix to forecast constraint on the neutrino mass.
\end{abstract}

%\pacs{95.36.+x, 98.80.Es, 95.35.+d}

\maketitle

\section{Introduction}

The evidence of neutrino oscillation implies that at least two neutrinos are massive \cite{ref:NOS1998}. The differences of neutrino masses squared in a standard scenario are known with three massive eigenstates \cite{ref:muMass}
\begin{eqnarray}
\Delta m^2_{12}=7.54^{+0.26}_{-0.22}\times 10^{-5} \text{eV}^2,\\
|\Delta m^2_{32}|=2.3^{+0.06}_{-0.06}\times 10^{-3} \text{eV}^2.
\end{eqnarray}   
But the mass hierarchy (the sign of $|\Delta m^2_{32}|$) is still difficult to know. From the above relations, one can easily derive the lower limit on the sum of neutrino mass $\sum_i m_{\nu,i}>0.057\text{eV}$. On the other hand, the measurement of the angular anisotropies of the comic microwave background (CMB) radiation puts the upper limit on the sum of neutrino mass $\sum_i m_{\nu,i}<0.23\text{eV}$ as reported by {\it Planck} 2015 \cite{ref:planck2015CP}, and future 21cm and precise CMB polarization observations \cite{ref:201621cm}.    

Massive neutrino cosmology has already been studied extensively in the literature (see \cite{ref:NCbook} for a comprehensive review). In the early Universe before the last scattering of the CMB photons, the neutrinos play the role as cosmic radiation because of their small total mass, affecting the matter-radiation equality time. This is the so-called early integrated Sacks-Wolfe (eISW) effect, which can be observed from the first peak position of CMB temperature anisotropic power spectrum. Subsequently in the matter and dark energy domination eras, neutrinos become non-relativistic and contribute to the total matter density in the Universe at the background and perturbation levels. Therefore, the geometric and dynamic measurements are useful to determine the neutrino mass. 

In the past few years, the linear perturbation theory with the presence of neutrino mass has been well understood \cite{ref:NCbook}. In the early Universe, when the neutrino behaves with relativistic degrees of freedom, its density fluctuation does not grow. Once neutrino thermal energy drops below its mass, it becomes non-relativistic at a redshift $z_{nr}$
\begin{equation}
z_{nr}(m_{\nu})=\frac{m_{\nu}}{5.28\times 10^{-4}\text{eV}}-1.
\end{equation}   
Thereafter, neutrino contributes to the total dark matter background density \cite{ref:Lesgourgues14}
\begin{equation}
\Omega_m=\Omega_c+\Omega_b+\Omega_{\nu},\quad \Omega_{\nu}=\frac{\sum_i m_{\nu,i}}{93.14h^2\text{eV}},
\end{equation}
where $h$ is related to the present Hubble parameter $H_0=100h \text{ km} \text{ s}^{-1}\text{ Mpc}^{-1}$. After the non-relativistic transition, the free streaming scale of neutrino changes from the Hubble scale to 
\begin{equation}
k_{fs}=0.776(1+z)^{-2}\frac{H(z)}{H_0}\left(\frac{m_{\nu}}{1\text{eV}}\right)h \text{Mpc}^{-1},
\end{equation}
which has a minimum at $z=z_{nr}$
\begin{equation}
k_{nr}=k_{fs}(z_{nr})\approx 0.0178\Omega^{1/2}_{m} \left(\frac{m_{\nu}}{1\text{eV}}\right)^{1/2}h \text{Mpc}^{-1},
\end{equation}
in the matter domination, considering $H(z)/H_0\approx \Omega^{1/2}_{m}(1+z)^{3/2}$ \cite{ref:NCbook}. Here, the scale $k_{nr}$ is the largest scale that can be affected by the presence of neutrino fluctuation. On smaller scales $k>k_{nr}$, density fluctuations are washed out, while on larger scales $k<k_{nr}$ neutrino behaves as cold dark matter. Thus, at sufficiently small scales $k\gg k_{nr}$, the power spectrum of matter $P(k)$ is depressed due to the lack of neutrino power. This is due to the modification of the linear evolution of density perturbation at small scales $k\gg k_{nr}$ by the presence of the neutrino \cite{ref:Angeliki}
\begin{equation}
\delta''+2\mathcal{H}\delta'=4\pi G \rho a^2(1-f_{\nu})\delta,\label{eq:PerNeutrino}
\end{equation}
where $\delta=\delta\rho_m/\rho_m$ with $\rho_m$ and $\delta \rho_{m}$ being the density and the overdensity of matter respectively; the prime denotes the derivative with respect to the conformal time $\tau$, and $f_{\nu}$ reads as \cite{ref:Shoji}
\begin{equation}
f_{\nu}\equiv\frac{\Omega_{\nu}h^2}{\Omega_{m}h^2}=\frac{1}{\Omega_{m}h^2}\frac{\sum_i m_{\nu,i}}{94.14\text{eV}}.\label{eq:fnu}
\end{equation}
 In an Einstein-de Sitter universe, one has a simple solution of the Eq. (\ref{eq:PerNeutrino})
\begin{equation}
\delta \propto a^{p},
\end{equation}
where $p\approx 1-3f_{\nu}/5$. Thus the suppression of the matter power spectrum is crudely estimated to be $\Delta P/P\approx -8f_{\nu}$ \cite{ref:HuNeutrino}. As expected, this suppression will change the relation between the potentials and matter density contrast at small scales $k\gg k_{nr}$ in the presence of neutrino, as
\begin{equation}
\nabla^2[\Phi(\vec{x},\tau)-\Psi(\vec{x},\tau)]=-8\pi G a^2\bar{\rho}(1-f_{\nu})\delta(\vec{x},\tau),
\end{equation}
where $\Phi$ is the Bardeen's curvature perturbation during the matter-dominated era and is related to the trace of metric as $g_{ii}=3a^2(1+2\Phi)$, whereas $\Psi$ is given by the component $g_{00}=-a^2(1+2\Psi)$. In the absence of significant sources of anisotropic stress, one gets $\Phi=-\Psi$. In fact, Eq. (\ref{eq:PerNeutrino}) follows from the above Poisson equation with vanishing anisotropic stress. In the linear perturbation theory, the evolution of the density fluctuation does not depend on the scales, so that the density fluctuation can be factorized as $\delta(\vec{x},\tau)=D(\tau)\delta(\vec{x})$, where $\delta(\vec{x})$ is the initial density fluctuation. Then the perturbation equation (\ref{eq:PerNeutrino}) becomes  
\begin{eqnarray}
D''&+&2\mathcal{H}D'+\frac{3}{2}\Omega_{m0}\mathcal{H}^2_0(1-f_{\nu})D=0,\label{eq:Dnu}\\
\frac{d f}{d\ln a}&+&f^2+(1-\frac{d\ln \mathcal{H}}{d\ln a})f=\frac{3}{2}(1-f_{\nu})\Omega_{m}(a),\label{eq:growthfnu}
\end{eqnarray}
at small scales $k\gg k_{nr}$ in the presence of neutrino, where $f=\frac{d\ln D}{d\ln a}$ is the growth factor. The suppression of the matter power spectrum at scales $k\gg k_{nr}$ in the presence of neutrino changes the depth of gravitational potential in the path of the CMB photons propagating from the last scattering surface to us. This modification finds signature in the anisotropies of the CMB photons caused by the gravitational anisotropies, which are observed as the gravitational lensing, the integrated Sachs-Wolfe (ISW) effects and its non-linear extension, Rees-Sciama (RS) effect \cite{ref:HuDold}. Of course, the scattering secondaries (including the thermal and kinetic Sunyaev-Zel'dovich (t/kSZ) effect) will also enter into the final anisotropies of the CMB photons \cite{ref:HuDold}. But in this work we will mainly focus on RS effect due to its sensitivity to the neutrino mass. And the tSZ effect can be removed because of its frequency dependence of the photon intensities. 

Below an arc minute scale, the temperature fluctuation caused by RS effect is of the order $\Delta T^{RS}/T\sim 10^{-8}$ \cite{ref:RSseljak,ref:RSTuluie}, which is smaller than that of primary CMB or that of thermal SZ effect by order of magnitude three \cite{ref:Atsushithesis}. For extracting this tiny fluctuation of RS effect in the total CMB temperature fluctuations, it should be correlated with the matter distribution of large scale structure (LSS), because the RS effect is generated by the LSS. 
The distribution of matter could be measured by the distribution of galaxy distribution and statistics of weak lensing. However, the galaxy distribution has crucial disadvantage due to the bias problem, $\delta_{g}=b\delta_{dm}$, i.e., the density contrast of galaxy is proportional to that of dark matter \cite{ref:Atsushithesis}. On large scales where the density contrast is small $\delta<1$, this relation is viable, but on small scales the relation is not valid and the bias largely depends on the survey, thus on the galaxy population and the scale $k$, say $P_{g}(k)=b^2\frac{1+Qk^2}{1+Ak}P_{\rm{lin}}(k)$ for instance \cite{ref:Qmodel}. Thus the galaxy distribution will not be reliable or robust tracer for the matter distribution. For instance, the convergence of weak lensing never suffers from the bias problem \cite{ref:Atsushithesis}. 

Fortunately, the cross-correlation between RS-$\kappa$ (the convergence of lensing) is about $\Delta T/T\simeq 10^{-14}$, which is much larger than that of kSZ effect \cite{ref:kSZconvergence,ref:RSkappa}. As far as the non-linear extension of ISW effect, i.e RS effect, is concerned, one has to understand the non-linear matter power spectrum, which is the incarnation of the distribution of matter at small scales. Usually, the non-linear matter power spectrum can be obtained from the N-body simulation, the halo model \cite{ref:halomodel}, or the standard perturbation theory (SPT) \cite{ref:SPT}. We will mainly focus on the third order SPT theory, because it can easily be extended to various cosmological models as compared to the N-body simulation and halo model, and also provides exact calculation in the quasi-linear regime \cite{ref:Lee2014,ref:Lee2015}. But it should be noted that the main conclusion obtained in this work does not depend on any nonlinear power spectrum method, because the sign flipping of cross-correlation power spectrum $C_{l}^{\rm{RS}-\kappa}$ is a common feature of the non-linear evolution of the density contrast at the smaller scales. Of course, one expects that different method will give almost the same nonlinear power spectrum, if the methods are consistent. This is also the other main reason to use SPT in this work.    

The remaining part of this paper is structured as follows. In section \ref{sec:CCRSSPT}, we give a brief introduction to the cross correlation between RS-$\kappa$ (the convergence of lensing) and standard perturbation theory (SPT), where the non-linear matter matter power spectrum at the third order will be presented. In section \ref{sec:SFSNR}, the dependence of the sign flipping of $C_{l}^{\rm{RS}-\kappa}$ on $\Omega_\nu$ is given. The signal to noise ratio based on deep and shallow survey is calculated. Section \ref{sec:conclusion} presents the conclusion.

 \section{Cross Correlation between RS-$\kappa$ (the convergence of lensing) and SPT} \label{sec:CCRSSPT}
 
 Here we just present the formulas that would be useful in this work. Detailed derivation of the correlation between RS effect and convergence $\kappa$ can be found in \cite{ref:Atsushithesis} and \cite{ref:RSkappa}. In Ref. \cite{ref:RSkappa} , the influence of dark energy to the cross correlation was investigated, for a given weak-lensing survey or the radial distribution of source galaxies $n(z)$ \cite{ref:Efstathiou,ref:Semboloni,ref:Hennawi}, 
\begin{equation} 
n(z) = A z^2 \exp[-(z/z_0)^{\beta}], \label{nz}
\end{equation} 
where the normalization factor $A$ is determined by $\int_{0}^{\infty} n(z) dz = 1$ (see Ref. \cite{ref:Husl} for other form). The total cross correlation angular power spectrum, $C_{l}^{\rm{RS}-\kappa}$ reads as
\begin{equation}
C_{l}^{\rm{RS}-\kappa} = \int_{0}^{z_s} dz_s n(z_s) C_{l}^{\rm{RS}-\kappa}(z_s), \label{Cl} 
\end{equation} 
where the cross-correlation power spectrum $C_{l}^{\rm{RS}-\kappa}(z_s)$ is calculated by \cite{ref:Atsushithesis}, see also in \cite{ref:RSkappa}
\begin{eqnarray}
C_{l}^{\rm{RS}-\kappa}(z_s)&=& \frac{4}{\pi} \int dk k^4 \int_{0}^{r_{\ast}} dr \int_{0}^{r_{s}} dr' \frac{r'(r_s - r')}{r_s}  P_{\Phi\Phi'}(k;r,r') j_{l}(kr) j_{l}(kr'), \nonumber\\
&=&2l^2  \int_{0}^{r_{s}} dr  \frac{r_s - r}{r^3 r_s}  P_{\Phi\Phi'}(k;r)|_{k=l/r}\label{ClRSkappazs}.
\end{eqnarray}
Here $j_{l}$ is a spherical Bessel function, and $r$ is the comoving distance at $z$. The second line of the above equation is obtained through the Limber's approximation \cite{ref:Limber}. And the power spectrum of the gravitational potential and its time derivative, $P_{\Phi\Phi'}$ are given by \cite{ref:Atsushithesis}, see also in \cite{ref:RSkappa}
\begin{equation}
P_{\Phi \Phi'} (k,\tau)=  \left(\frac{3}{2}\frac{\Omega_{\rm m0}H^2_0}{c^2ak^2} \right)^2 \left[ P_{\delta \delta'}(k,\tau) - \mathcal{H} P_{\delta\delta}(k,\tau) \right] \label{PPhipPhi2}. 
\end{equation} 
In the linear regime, where $\delta'(z)=D'(z)\delta(0)=f\mathcal{H}\delta(z)$ is respected, the cross power spectrum is given as 
\begin{equation}
P_{\Phi \Phi'} (k,\tau)=  \left(\frac{3}{2}\frac{\Omega_{\rm m0}H^2_0}{c^2ak^2} \right)^2 \left[ \mathcal{H}(z)\left\{ f(z)-1 \right\}  \right]P^{\rm{lin}}_{\delta\delta}(k,\tau).
\end{equation}
Since $f(z)$ is in the range of $[0,1]$, the cross power spectrum $P_{\Phi \Phi'} (k,\tau)$ will not outdo zero. However, we need the cross correlation angular power spectrum below arc minute scale, where the power spectrum requires the non-linear treatment. For doing that, the third order SPT should be employed. In this work, we will adopt two models as samples, (Model I) Deep Survey, $(\beta, z_0)=(0.75,0.5)$, whose $n(z)$ peaks at $z\sim 2.2$ with a broad distribution, and (Model II) Shallow Survey, $(\beta, z_0)=(2,0.9)$, which peaks at $z\sim 0.9$ with a narrow distribution. These deep and shallow surveys were also studied in Ref. \cite{ref:Atsushithesis}, see also in \cite{ref:RSkappa}

Now, we present the main results of the matter power spectrum based on SPT. Considering the matter as continuous fluid, its evolution is governed by the continuity and Euler equations,
\begin{eqnarray}
\delta'+\nabla\cdot [\vec{v}(1+\delta)]&=&0,\\
\vec{v}'+2\mathcal{H}\vec{v}+(\vec{v}\cdot\nabla)\vec{v}&=&\nabla\Phi,
\end{eqnarray}
where $\vec{v}=\partial \vec{x}/\partial \tau$ is the conformal velocity.  As a natural extension of linear perturbation theory, the density fluctuation and divergence of velocity are expanded in series as \cite{ref:SPT}
\begin{eqnarray}
\delta(\vec{k},\tau)&=&\sum^{\infty}_{n=1}\delta_{n}(\vec{k},\tau),\\
\theta(\vec{k},\tau)&=&\sum^{\infty}_{n=1}\theta_{n}(\vec{k},\tau),
\end{eqnarray} 
where $\theta=\vec{k}\cdot \vec{v}$ is the divergence of the velocity. In a cosmological model, for example $\Lambda$CDM model, the above series expansion can be written as \cite{ref:SPT}
\begin{eqnarray}
\delta(\vec{k},\tau)&=&\sum^{\infty}_{n=1}D^{n}(\tau)\delta_{n}(\vec{k}),\\
\theta(\vec{k},\tau)&=&\sum^{\infty}_{n=1}D'D^{n-1}(\tau)\theta_{n}(\vec{k}).
\end{eqnarray}  
These are simple extensions from EdS to a cosmological model, where $n$-th variables $\delta_{n}$ and $\theta_{n}$ read as \cite{ref:SPT}
\begin{eqnarray}
\delta_{n}(\vec{k})&=&\int \frac{d^3 q_{1}}{(2\pi)^3}\cdot\cdot\cdot\frac{d^3 q_{n}}{(2\pi)^3}\delta_{D}\left(\sum^{n}_{i=1}\vec{q}_i-\vec{k}\right)F^{(s)}_{n}(\vec{q}_i,\cdot\cdot\cdot\vec{q}_n)\delta_{1}(\vec{q}_1)\cdot\cdot\cdot\delta_{1}(\vec{q}_n),\\
\theta_{n}(\vec{k})&=&\int \frac{d^3 q_{1}}{(2\pi)^3}\cdot\cdot\cdot\frac{d^3 q_{n}}{(2\pi)^3}\delta_{D}\left(\sum^{n}_{i=1}\vec{q}_i-\vec{k}\right)G^{(s)}_{n}(\vec{q}_i,\cdot\cdot\cdot\vec{q}_n)\delta_{1}(\vec{q}_1)\cdot\cdot\cdot\delta_{1}(\vec{q}_n),
\end{eqnarray}
where $F^{(s)}_{n}$ and $G^{(s)}_{n}$ are symmetric mode coupling kernels. Here $F^{(s)}_{1}=1$ and $G^{(s)}_{1}=1$, and for the $\Lambda$CDM model, the second order variables are given by \cite{ref:SPT}
\begin{eqnarray}
F^{(s)}_{2}(\vec{k}_1,\vec{k}_2)&=&\frac{5}{7}+\frac{2}{7}\frac{(\vec{k}_1\cdot \vec{k}_2)^2}{k^2_1 k^2_2}+\frac{\vec{k}_1\cdot \vec{k}_2}{2}\left(\frac{1}{k^2_1}+\frac{1}{k^2_2}\right),\\
G^{(s)}_{2}(\vec{k}_1,\vec{k}_2)&=&\frac{3}{7}+\frac{4}{7}\frac{(\vec{k}_1\cdot \vec{k}_2)^2}{k^2_1 k^2_2}+\frac{\vec{k}_1\cdot \vec{k}_2}{2}\left(\frac{1}{k^2_1}+\frac{1}{k^2_2}\right).
\end{eqnarray}
It should be noted that the above expansion is not exactly correct even in the $\Lambda$CDM model. However, it was reported that the largest deviation in the density perturbation variables from the cosmology is almost entirely encoded into the linear growth factor, and the contribution from other terms can be less than one percent \cite{ref:Jeong}. The gaussianity of $\delta_1$ suggests that all the odd order moments vanishes. Therefore, the matter power spectrum reads as
\begin{equation}
P(k,\tau)=D^2(\tau)P^{11}_{\delta\delta}(k)+D^4(\tau)\left[P^{22}_{\delta\delta}(k)+2P^{13}_{\delta\delta}(k)\right],\label{eq:Pks}
\end{equation}   
where $P^{11}_{\delta\delta}(k)$ is the linear power spectrum, and $P^{22}_{\delta\delta}(k)$ and $P^{13}_{\delta\delta}(k)$ are quartic parts for $\delta_{1}$ \cite{ref:Atsushithesis},
\begin{eqnarray}
P^{22}_{\delta\delta}(k) &\equiv& \langle \delta_{2}(\vec{k})\delta^{\ast}_{2}(\vec{k})\rangle \nonumber\\
&=&2\int \frac{d^3 q}{(2\pi)^3} P^{11}_{\delta\delta}(q)P^{11}_{\delta\delta}(|\vec{k}-\vec{q}|)[F^{(s)}_2(\vec{q},\vec{k}-\vec{q})]^2,\\
2P^{13}_{\delta\delta}(k) &\equiv& \langle \delta_{1}(\vec{k})\delta^{\ast}_{3}(\vec{k})\rangle \nonumber\\
&=&6P^{11}_{\delta\delta}(k)\int \frac{d^3 q}{(2\pi)^3} P^{11}_{\delta\delta}(q)F^{(s)}_3(\vec{q},-\vec{k},\vec{k}).
\end{eqnarray}

\begin{center}
\begin{figure}[tbh]
\includegraphics[width=9.5cm]{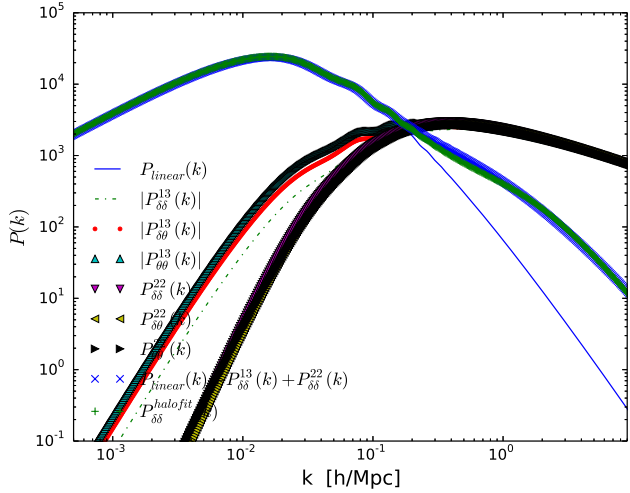}
\caption{The power spectra at redshift $z=0$ obtained from the third order standard perturbation theory and {\bf  halofit} \cite{ref:HALOFIT}.}\label{fig:grhalopks}
\end{figure}
\end{center}

For illustration, we plot the linear and non-linear power spectra at the redshift $z=0$ with $\Omega_{\nu}=0$ in Figure \ref{fig:grhalopks}, where the linear matter power spectrum is calculated by {\bf CAMB} \cite{ref:CAMB}. The non-linear power spectrum is calculated through the Eq. (\ref{eq:Pks}) and the {\bf  halofit} \cite{ref:HALOFIT}. As seen in Figure \ref{fig:grhalopks}, we also show the second and third order components such as $P^{13}_{\delta\delta}\equiv \langle \delta_{1}(\vec{k})\delta^{\ast}_{3}(\vec{k})\rangle$, $P^{13}_{\delta\theta}\equiv \langle \delta_{1}(\vec{k})\theta^{\ast}_{3}(\vec{k})\rangle$, $P^{13}_{\theta\theta}\equiv \langle \theta_{1}(\vec{k})\theta^{\ast}_{3}(\vec{k})\rangle$, $P^{22}_{\delta\delta}\equiv \langle \delta_{2}(\vec{k})\delta^{\ast}_{2}(\vec{k})\rangle$, $P^{22}_{\delta\theta}\equiv \langle \delta_{2}(\vec{k})\theta^{\ast}_{2}(\vec{k})\rangle$ and $P^{22}_{\theta\theta}\equiv \langle \theta_{2}(\vec{k})\theta^{\ast}_{2}(\vec{k})\rangle$.
%\begin{widetext}
%\end{widetext}

\begin{center}
\begin{figure}[tbh]
\includegraphics[width=9.5cm]{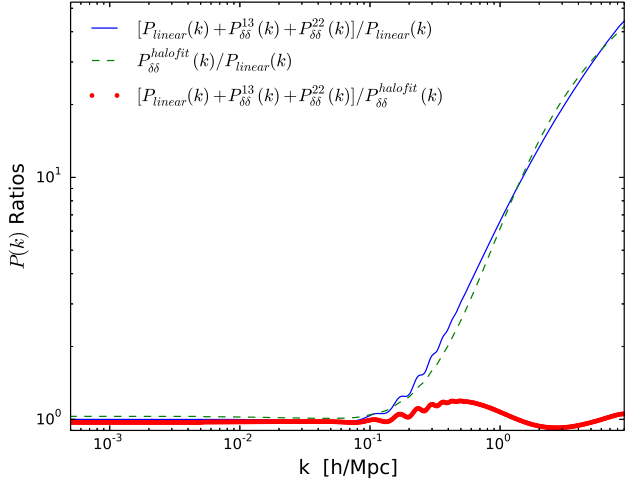}
\caption{The ratio of the matter power spectrum at redshift $z=0$ with the linear one for the {\bf  halofit} \cite{ref:HALOFIT} and third order standard perturbation theory.}\label{fig:grhalopksratio}
\end{figure}
\end{center}

For cross-checking and comparing the non-linear matter power spectrum obtained from the {\bf  halofit} \cite{ref:HALOFIT}, the third order standard perturbation theory and the linear matter power spectrum, we plot the ratios in Figure \ref{fig:grhalopksratio} for the $\Lambda$CDM model, where the cosmological model parameters are fixed to their values shown in {\it Planck}2015 results \cite{ref:planck2015MG}. One can clearly see that the non-linear matter power spectrum obtained from third order standard perturbation theory can match very well with the one obtained from the {\bf  halofit} \cite{ref:HALOFIT}. 
%\begin{widetext}
%\end{widetext}

\section{Sign Flipping of $C_{l}^{\rm{RS}-\kappa}$ on $\Omega_\nu$ and the Signal to Noise Ratio} \label{sec:SFSNR}

\begin{center}
\begin{figure}[tbh]
\includegraphics[width=9.5cm]{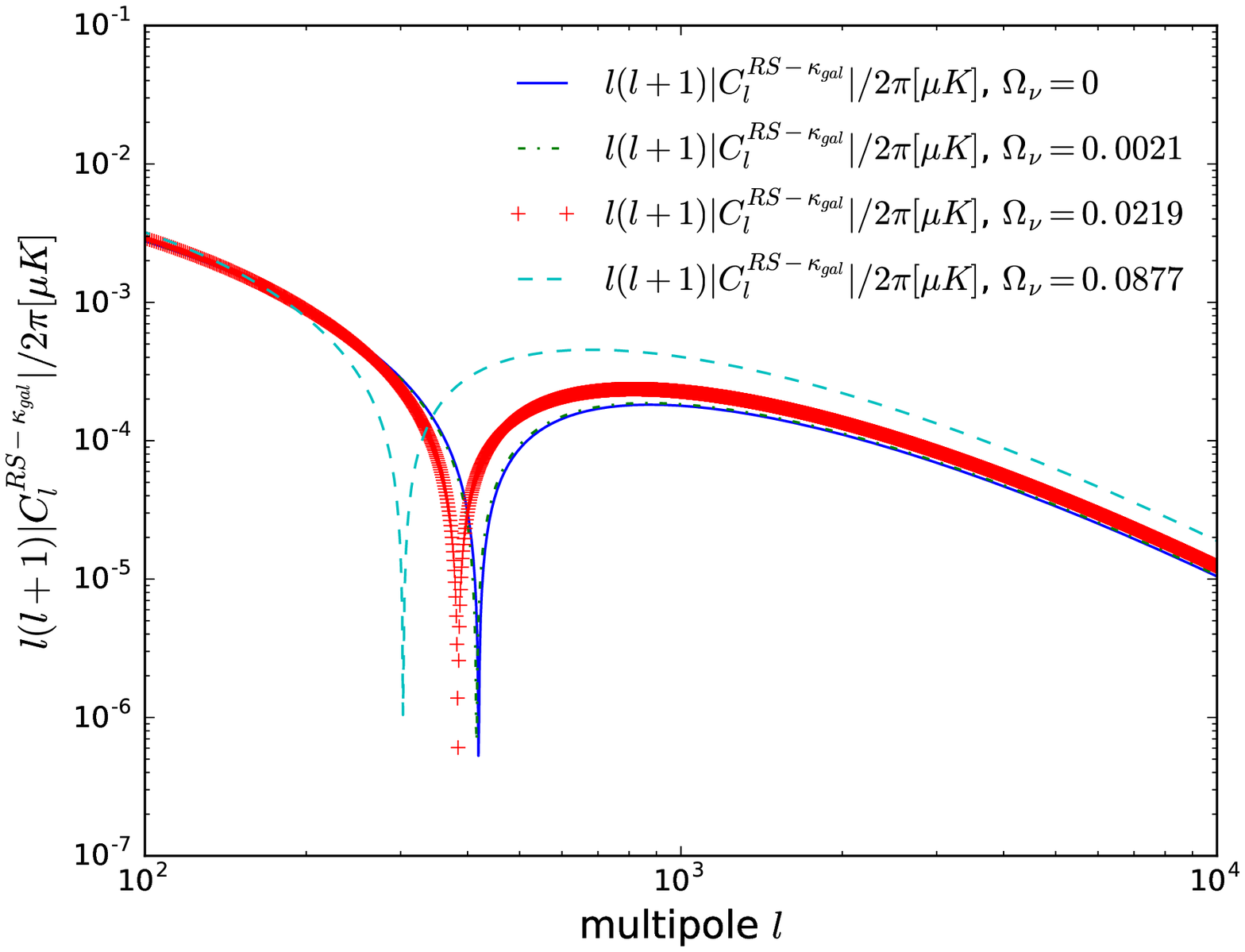}
\includegraphics[width=9.5cm]{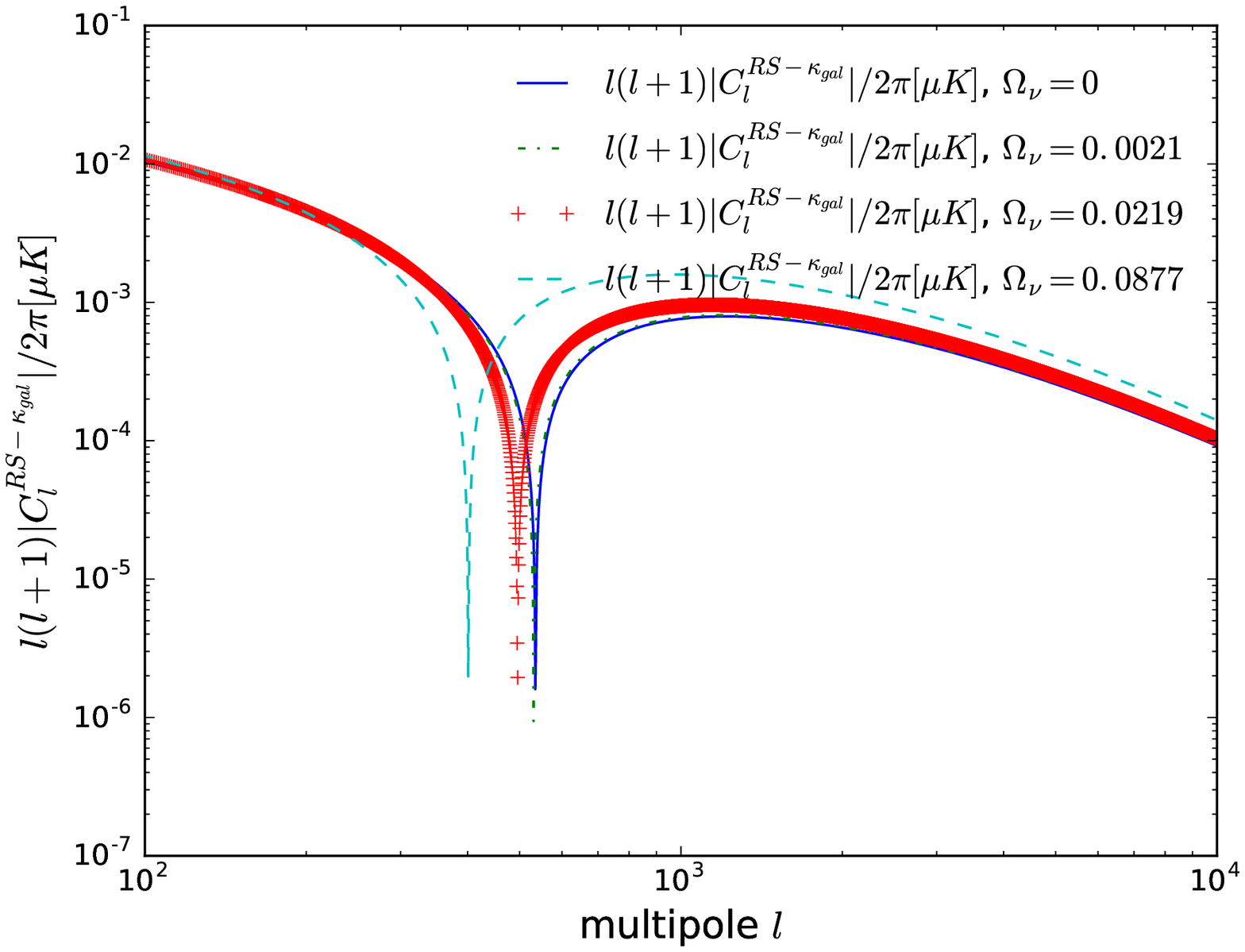}
\caption{$C_{l}^{\rm{RS}-\kappa}$ for Shallow Survey (the upper panel) and Deep Survey (the lower panel) with different values of $\Omega_{\nu}$.}\label{fig:cls}
\end{figure}
\end{center}

The power spectrum $P_{\delta\delta'}$ can be calculated as
\begin{equation}
P_{\delta\delta'}=\frac{1}{2}\frac{\partial}{\partial \tau}P_{\delta\delta}.
\end{equation}
Using Eq. (\ref{eq:Pks}), one has
\begin{eqnarray}
P_{\delta\delta'}&=&D'DP^{11}_{\delta\delta}(k,0)+2D'D^3\left[P^{22}_{\delta\delta}(k,0)+2P^{13}_{\delta\delta}(k,0)\right]\nonumber\\
&=&f\mathcal{H}P^{11}_{\delta\delta}(k,z)+2f\mathcal{H}\left[P^{22}_{\delta\delta}(k,z)+2P^{13}_{\delta\delta}(k,z)\right].
\end{eqnarray}
Thus, Eq. (\ref{PPhipPhi2}) can be rewritten as
\begin{eqnarray}
P_{\Phi \Phi'} (k,\tau) &=& \left(\frac{3}{2}\frac{\Omega_{\rm m0}H^2_0}{c^2ak^2} \right)^2 \left[ P_{\delta \delta'}(k,\tau) - \mathcal{H} P_{\delta\delta}(k,\tau) \right],\nonumber\\
&=&\mathcal{H}(f-1)P^{11}_{\delta\delta}(k,z)+\mathcal{H}(2f-1)\left[P^{22}_{\delta\delta}(k,z)+2P^{13}_{\delta\delta}(k,z)\right].
\end{eqnarray}
From the above equation, one can easily find that the power spectrum $P_{\Phi \Phi'} (k,\tau)$ cannot surpass zero in the linear regime due to the fact that $f\le 1$. And there will be a sign change of the power spectrum $P_{\Phi \Phi'} (k,\tau)$. This sign-flipping is caused by the sign change of \cite{ref:Lee2015}
\begin{equation}
\Phi' \propto \mathcal{H}\delta\left(\frac{d\ln \delta}{d\ln a}-1\right)=\mathcal{H}\delta\frac{d\ln (\delta/a)}{d\ln a}.
\end{equation}  
Thus if the density contrast grows faster than the background scale factor $\delta\sim a$, there would be a sign flipping. And the flipping scale will depend on the growth history for a cosmological model and this situation happens only in the non-linear regime. The dependence of the flipping scale of the cross power spectrum for $C_{l}^{\rm{RS}-\kappa}$ on the neutrino mass, $\Omega_\nu$, can be easily seen in Figure \ref{fig:cls}, where the shallow (the left panel) and deep survey (the right panel) are employed as examples. For the Shallow survey, the flipping scales are  $\ell=303$ ($\Omega_\nu=0.0877$), $\ell=383$ ($\Omega_\nu=0.0219$) and $\ell=415$ ($\Omega_\nu=0.0021$). For the Deep survey, the flipping scales are  $\ell=401$ ($\Omega_\nu=0.0877$), $\ell=495$ ($\Omega_\nu=0.0219$) and $\ell=531$ ($\Omega_\nu=0.0021$). One reads from the cross power spectrum for $C_{l}^{\rm{RS}-\kappa}$ that larger values of $\Omega_\nu$ make the flipping scale at lower multipole $\ell$. One can also see that the flipping scale depends on the galaxy source distribution function due to the different peak position of different surveys.    
%\begin{widetext}
%\end{widetext}

\begin{center}
\begin{figure}[tbh]
\includegraphics[width=9.5cm]{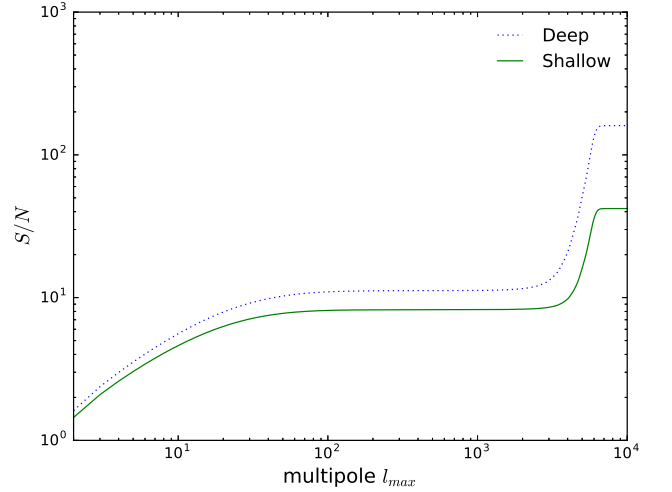}
\caption{The cumulative signal-to-noise with respect to the multipole $\ell_{max}$ for the Shallow (the solid line) and Deep (the dotted line) surveys.}\label{fig:sn}
\end{figure}
\end{center}

One would also like to see whether the correlation can be detected in the future observations. The significance of detection is quantified by the signal-to-noise ratio ($S/N$), which at each multipole $\ell$ is defined as \cite{ref:Atsushithesis}
\begin{equation}
\left(\frac{S}{N}\right)^2_{\ell}\approx f_{sky}Cov^{-1}_{\ell}(C^{RS-\kappa}_{\ell})^2,
\end{equation}
where $f_{sky}$ is the fraction of the sky where both CMB and WL are observed. Further, $Cov_{\ell}$ is the covariance matrix defined by  \cite{ref:Atsushithesis}
\begin{equation}
Cov_{\ell}=\frac{\tilde{C}^{CMB}_{\ell}\tilde{C}^{\kappa}_{\ell}+(\tilde{C}^{RS-\kappa}_{\ell})^2}{2\ell+1},\label{eq:covmatrix}
\end{equation}
where $\tilde{C}=C_{\ell}+N_{\ell}$ is a summation of the true angular auto/correlation power spectrum and a noise spectrum. Then noise power spectrum for CMB and convergence $\kappa$ are defined as follows \cite{ref:Knox1995,ref:Schncider2005}
\begin{eqnarray}
N^{CMB}_{\ell}&=&\sigma^2_{pix}\theta^2_{fwhm}\exp[\ell(\ell+1)\theta^2_{fwhm}/8\ln 2],\\
N^{\kappa}_{\ell}&=&\sigma^2_{\gamma}/n_{gal},
\end{eqnarray}
where $\sigma_{pix}$ is the sensitivity to CMB temperature fluctuation in units of background temperature; $\theta_{fwhm}$ is the full width half maximum of the gaussian beam size; $\sigma_{\gamma}$ is the dispersion on the intrinsic ellipticities f the lensed galaxies, and $n_{gal}$ is the number density of galaxies of the lensing survey per unit steradian. In this paper, we adopt $\sigma_{pix}=4.3\times 10^{-6}$, $\theta_{fwhm}=5.5arcmin$, $f_{sky}=0.8$, $\sigma_{\gamma}=0.1$ and $n_{gal}=100/arcmin^2$ as an example which are survey parameters of the LSST and {\it Planck} \cite{ref:Atsushithesis}. The cumulative signal-to-noise ratio is written as the summation of the $S/N$ at each multipole  \cite{ref:Atsushithesis}
\begin{equation}
\left(S/N\right)^2=\sum^{\ell_{max}}_{\ell=\ell_{min}}\left(S/N\right)^2_{\ell}.
\end{equation} 
We show the cumulative $S/N$ as a function of $\ell_{max}$ in Figure \ref{fig:sn}, where $\ell_{min}=2$ is adopted. We found these $f_{sky}=0.8$ surveys can yield $S/N\sim 160~(42)$ for Deep (Shallow) WL surveys. And it is clear that the Deep surveys have large signal-to-noise ratio. 
%\begin{widetext}
%\end{widetext}

\begin{center}
\begin{figure}[tbh]
\includegraphics[width=9.5cm]{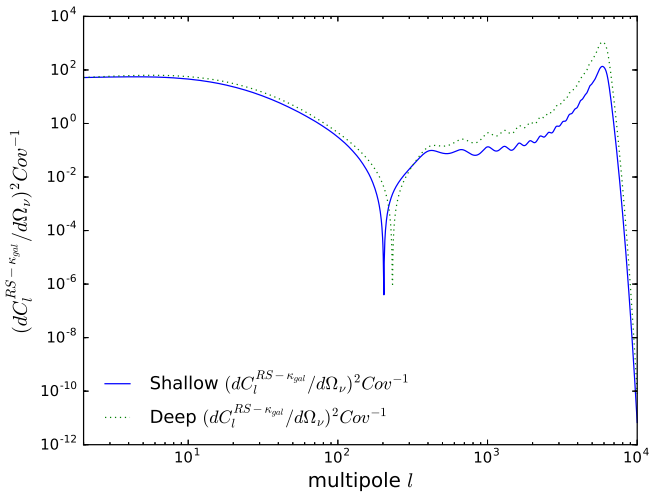}
\caption{The signal to noise ratio squared at each multipole $\ell$, $(d C^{RS-\kappa}_{\ell}/d\Omega_{\nu})^2 Cov^{-1}$ for the deep (green doted line) and shallow (blue solid line) surveys.}\label{fig:dcldomeganu}
\end{figure}
\end{center}

The Fisher information matrix is a useful tool to estimate the upper bound on the parameter error $\Delta$ on a parameter $\theta_{\alpha}$ according to the Cramer-Rao inequality,
\begin{equation}
\Delta\theta_{\alpha}\leq \sqrt{(F^{-1})_{\alpha\alpha}},
\end{equation} 
where $F_{\alpha\beta}$ is the Fisher information matrix which is defined as
\begin{equation}
F_{\alpha\beta}=\left\langle \frac{\partial^2\ln \mathcal{L}}{\partial \theta_{\alpha}\partial \theta_{\beta}}\right\rangle,
\end{equation}
where $\mathcal{L}$ is the likelihood function. Following Ref. \cite{ref:Atsushithesis}, for the cross correlation the Fisher matrix is given by
\begin{equation}
F_{\alpha\beta}=\sum^{\ell_{max}}_{\ell=\ell_{min}}\frac{\partial C^{RS-\kappa}_{\ell}}{\partial \theta_{\alpha}}Cov^{-1}_{\ell}\frac{\partial C^{RS-\kappa}_{\ell}}{\partial \theta_{\beta}},
\end{equation} 
where $Cov_{\ell}$ is the covariance matrix defined in Eq. (\ref{eq:covmatrix}), $\ell_{min}=2$ and $\ell_{max}=10^4$. In this work, we are interested to the cosmological parameters $\Omega_{\nu}=f_{\nu}\Omega_{m}$. Thus we take it as the only one variable parameter and fix the other relevant cosmological parameters to their best fit values obtained by {\it Planck}2015 \cite{ref:planck2015CP}. As studied in Ref. \cite{ref:Atsushithesis}, large values of $\partial C^{RS-\kappa}_{\ell}/\partial\theta_{\alpha}$ make large values of Fisher matrix. Thus a tight constraint to the model parameter an be obtained. In figure \ref{fig:dcldomeganu}, we show the signal to noise ratio squared at each multipole $\ell$, $(d C^{RS-\kappa}_{\ell}/d\Omega_{\nu})^2 Cov^{-1}$ for the Deep (green doted line) and Shallow (blue solid line) surveys respectively. One can see that the signal to noise ratio is suppressed due to the detector noise of CMB at $\ell \ge 5000$. After calculating the Fisher matrix, one obtains the upper bound for $\Omega_{\nu}$: $\Delta\Omega_{\nu} \le  0.00220$ for shallow survey and $\Delta\Omega_{\nu} \le  0.000926$ for deep survey, which correspond to the neutrino mass: $\Delta \sum_i m_{\nu,i} \le 0.21 h^2\text{eV}$ for Shallow survey and  $\Delta \sum_i m_{\nu,i} \le 0.086 h^2\text{eV}$ for Deep survey. 
%\begin{widetext}
%\end{widetext}

\section{Conclusion} \label{sec:conclusion} 

We have studied the possibility to use the cross correlation between the Rees-Sciama effect and weak lensing to probe the neutrino mass. After the non-relativistic transition, massive neutrino contributes to the total matter density in the Universe at the background and perturbation levels. Due to the free streaming, the fluctuation of neutrino is washed out at the small scales $k\gg k_{fs}$. This results in the suppression of the matter power spectrum at the small scales about $\Delta P/P\approx -8f_{\nu}$. And this discrepancy is amplified due to gravitational attraction at the non-linear scales. In other words, the neutrino mass affects the non-linear evolution of the density contrast at the smaller scales. That makes a sign flipping on $\Phi' \propto \mathcal{H}\delta d\ln (\delta/a)/d\ln a$ at different scales/redshifts due to the fast growth of the density contrast in comparison to the background scale factor $\delta\sim a$ in the non-linear regime. And the flipping scale depends on the neutrino mass. For detaching the tiny fluctuation of RS effect from the total CMB temperature fluctuations, it is correlated with the matter distribution of large scale structure via weak lensing with Shallow and Deep surveys. We find that the sign flipping in the cross power spectrum of the Rees-Sciama effect and weak lensing with Shallow and Deep surveys depend on the neutrino mass: the larger values of $\Omega_\nu$ make the flipping scale at lower multipole $\ell$ for the cross power spectrum for $C_{l}^{\rm{RS}-\kappa}$. And the Deep survey has larger signal-to-noise ratio. The findings of this study are expected to be fruitful in the probe of neutrino mass.

\acknowledgements{The author thanks an anonymous referee for helpful improvement of this paper. This work is supported in part by National Natural Science Foundation of China under Grant No. 11275035, Grant No. 11675032 (People's Republic of China), and supported by 'the Fundamental Research Funds for the Central Universities' under Grant No. DUT16LK31.}

\end{document}